# A Method for Detecting Abnormal Data of Network Nodes Based on Convolutional Neural Network

*Xianhao Shen[1,2], Yihao Zang[1], Qiong Gui[1]，Jinsheng Yi[1]，Shaohua Niu[2]*

[1]Guilin University of Technology

Guilin 541006，China

2120190631@glut.edu.cn;*Corresponding author:25337698@qq.com

[2]Beijing Institute of Technology

Beijing 100081，China

**Abstract:** Abnormal data detection is an important step to ensure the accuracy and reliability of node data in wireless sensor networks. In this paper, a data classification method based on convolutional neural network is proposed to solve the problem of data anomaly detection in wireless sensor networks. First, Normal data and abnormal data generated after injection fault are normalized and mapped to gray image as input data of the convolutional neural network. Then, based on the classical convolution neural network, three new convolutional neural network models are designed by designing the parameters of the convolutional layer and the fully connected layer. This model solves the problem that the performance of traditional detection algorithm is easily affected by relevant threshold through self-learning data characteristics of convolution layer. The experimental results show that this method has better detection performance and higher reliability.

**Key words:** data anomaly detection; convolutional neural network; injection fault;normalized

1.**Introduction**.Wireless sensor network (WSN) is a network composed of many tiny, cheap and low-power sensor nodes, which are deployed intensively or sparsely to collect information or monitor objects in the environment [1]. In wireless sensor networks, the data collected by nodes has a significant deviation from the actual data, which is called abnormal data [2]. Due to the limitations of the computing power, storage space, power supply and other aspects of sensor nodes, and vulnerable to external interference, some nodes in the sensor network will be abnormal, which will affect the authenticity and stability of the data collected by sensor nodes, and lead to abnormal data. As an important technology to ensure data quality in WSN, abnormal data detection technology of wireless sensor network can improve the accuracy and reliability of data in wireless sensor network effectively.

At present, the anomaly detection methods for wireless sensor networks are mainly divided into statistical method[3,4], clustering method[5,6] and classification based method[7-10]. For example, researchers proposed an anomaly detection method based on hypothetical mathematical statistical model and kernel density function[3].In addition, an abnormal data detection algorithm for wireless sensor networks based on variable width histogram is proposed[4]. This algorithm reduce data transmission volume furtherly and saving communication cost By collecting histogram information of data flow distribution in wireless sensor network and dynamically merging histogram interval which can change the width of original interval. However, the method based on statistics needs to know the prior information of the data set in advance to establish the statistical model. Meanwhile, in many practical cases, the statistical model of data is too difficult to establish and it also exists limitations. A recent study[5][6] proposed an anomaly detection algorithm based on K-means algorithm. The core idea of the algorithm is to calculate the distance between data and cluster center，Then, the abnormal data can be detected by dividing the data instances into close data clusters. The advantage of clustering method is that it does not need statistical model of data, but it is difficult to determine the size and number of clusters. In the work of [7]，the authors used three different machine learning methods to classify and compare abnormal data and normal data, and then detect abnormal data. However, the robustness of the algorithm needs to be improved. In the work of [8], the authors proposed an anomaly detection algorithm based on SVM. Firstly, the algorithm uses the training data set to learn a classification model, and then divides the data instances into the learned classes. When the data belongs to less class data or does not belong to any classification, it is considered as abnormal data, but the effect of anomaly detection is not ideal. In the work of [9], the authors proposed a support vector machine (SVM) anomaly detection algorithm based on deep belief network. Firstly, the algorithm uses the deep belief network to reduce the dimensionality of the high-dimensional data, and then uses the support vector machine combined with the sliding window model to realize the detection of the reduced data. However, some obscure abnormal data could not be distinguished from normal data after dimensionality reduction Low detection accuracy. In the work of [10] uses classification and regression tree to detect abnormal data, but this method requires a lot of preprocessing for data with time series, which not only increases the extra cost of classification algorithm, but also reduces the accuracy of classification. The method based on classification is not limited by the distribution of data. It divides the data into normal data and abnormal data to distinguish abnormal data. However, the traditional classifier is difficult to capture all the features of abnormal data. In recent years, as a classification method in the field of machine learning, deep learning has made brilliant achievements in many fields with its powerful automatic feature extraction ability. Deep learning can solve the problems of feature extraction difficulty effectively, and strong subjectivity of threshold selection, meanwhile, it also can meet the requirements of massive data mining. In the field of anomaly detection, we can further improve the ability of anomaly detection with the unique performance of deep learning.

2. **An improved method for abnormal data detection based on convolutional neural network**

There are four key steps in anomaly detection of wireless sensor network data using convolutional neural network: firstly, preprocess the data set; secondly, design the appropriate convolutional neural network model; then, based on the proposed convolutional neural network model, we can learn the features of the preprocessed data. Finally, we test it on the trained model and complete the detection of abnormal data.

2.1 **Data set construction**. This paper uses the IBRL [13] (Intel Berkeley Laboratory) data set, which is collected by 54 sensor nodes arranged in Intel Berkeley Laboratory. The sampling interval is 31 seconds. The data of temperature, humidity, light intensity and voltage are collected at the same time. Deep learning needs a large number of data samples as the training set. However, the actual data collected by sensors is lack of a large number of abnormal data, so it is necessary to inject faults into normal data artificially. Through the analysis of the real data collected by the sensor, fixed fault, noise fault and short-term fault are three common sensor fault types that cause abnormal sensor data [14,15]. The main fault causes and fault injection methods are shown in Table.

TABLE 1. Sensor fault reason and fault injection method

| fault type | Cause of fault | Fault injection method |
| --- | --- | --- |
| Noise fault | Sensor hardware failure, battery power shortage and other reasons | $S_w[i] = S_W[i] + \varepsilon$ |
| Short-term fault | Abnormal battery voltage, sensor hardware connection failure and so on | $S[i][j] = S[i][j] + f * S[i][j]$ |
| Fixed fault | Sensor power supply failure, open circuit or short circuit | $S_w[i] = G$ |

Since the storage capacity of sensor nodes is limited and cannot meet the unlimited growth of time series data, this paper uses sliding windows to process data. The sliding window model uses a sliding window with length L (L > 0) to segment the sensor data stream into in window data and out window data, and there are L sampling data in the window. When the window slides, L samples in the window exit the window, and the L data at the next sampling time enter the window. $S_w[i]$ represents a group of consecutive raw data with length w randomly obtained from the sliding window data of the ith node. In this experiment, the sliding window length L is 64, W is 20, that is, the number of fault data is 20. Amog it, $\varepsilon = N(0, \sigma^2)$ follows the normal distribution, the standard deviation of noise $\varepsilon$ is r times that of normal data, the injection of noise fault is realized by adding all the data in $S_w[i]$ with the random number of normal distribution. $S[i][j]$ denotes the jth data point randomly selected by the ith node in the sliding window. Short time fault injection is realized by increasing its amplitude by f times, where f is a constant. Fixed fault injection is to set all the data in $S_w[i]$ as a fixed value G. in real life, the fixed value G is far greater than the data value collected by sensors under normal conditions.

Injecting fault into normal data set can control the abnormal strength of abnormal data and explore the performance limit of the detection method in this paper, thus it fully prove the adaptability and accuracy of the detection method in this paper. By setting the size of r to control the abnormal intensity of noise data, the size of F to control the abnormal intensity of short-term data, and the size of G to control the abnormal intensity of fixed data. In this paper, r is set to 0.5, 1, 1.5, 2, 2.5, 3, f is set to 1.5, 2, 3, 5, 7, 10, and G is set to 150, 300,500. Many anomalies observed in the existing actual data sets have relatively high anomaly intensity, and the anomaly intensity in future collected data sets is not necessarily the same as that in this paper. Therefore, it is very significant to understand the impact of a series of abnormal intensities on the performance of detection methods deeply.

In this paper, the single fault and mixed fault (two types of faults at the same time) occur in a single sensor node. The mixed fault needs to inject all the data in $S_w[i]$ into two kinds of faults successively. Mixed fault is an experiment based on a single fault. The purpose of this paper is to evaluate the detection performance of the algorithm in complex situations.

2.2. **Data preprocessing**. Convolutional neural network has a good effect in the processing of two-dimensional images, so it is necessary to preprocess the data set to form a two-dimensional matrix, and then convert the two-dimensional matrix into a gray-scale image. Most of the previous studies only focused on the single type of data of sensor nodes (such as only focusing on the characteristic data of temperature), but ignored the correlation between multi-dimensional data. For example, when the temperature of a node increases significantly, the humidity of the node will also decrease. Reference [12] studies the correlation between multi-dimensional data in the same node. This paper takes this as the theoretical basis and uses the correlation between multi-dimensional data to study the single fault and mixed fault of temperature data in IBRL data set.

The four features have different dimensions and dimension units. In order to eliminate the dimensional influence between them and in order to make the data comparable, this paper makes numerical normalization for each type of data in the data set. preprocessing the data adopt the deviation standardization method in this paper, and the original value $x$ is transformed into the preprocessed data $x'$. formula (1) can express $x'$ as follows:

$$X' = \frac{X - \min}{\max - \min} \tag{1}$$

Max and min represent the maximum and minimum values in the original data, respectively.

The normalized data needs to be converted into gray value Y, and the range of $Y$ is [0,255]. 0 means black, 255 means white, and the intermediate values are various gray tones from black to white. It can be expressed by formula (2) as follows:

$$Y = X' * 255 \tag{2}$$

According to the number of features in the data set, it is necessary to construct a matrix dimension suitable for convolution neural network learning. In addition, on the basic of the four characteristics of IBRL data set, 64 consecutive data of the four characteristic data in the same period are taken from the sliding window to construct a 16 16 16 matrix. formula (3) can express The matrix A as follows:

$$A = \begin{bmatrix} T_m & T_{m+1} & T_{m+2} & \cdots & T_{m+15} \\ H_m & H_{m+1} & H_{m+2} & \cdots & H_{m+15} \\ L_m & L_{m+1} & L_{m+2} & \cdots & L_{m+15} \\ V_m & V_{m+1} & V_{m+2} & \cdots & V_{m+15} \\ T_{m+16} & T_{m+17} & T_{m+18} & \cdots & T_{m+31} \\ H_{m+16} & H_{m+17} & H_{m+18} & \cdots & H_{m+31} \\ L_{m+16} & L_{m+17} & L_{m+18} & \cdots & L_{m+31} \\ V_{m+16} & V_{m+17} & V_{m+18} & \cdots & V_{m+31} \\ \cdots & \cdots & \cdots & \cdots & \cdots \\ L_{m+48} & L_{m+49} & L_{m+50} & \cdots & L_{m+63} \\ V_{m+48} & V_{m+49} & V_{m+50} & \cdots & V_{m+63} \end{bmatrix} \tag{3}$$

$T_m$、$H_m$、$L_m$ and $V_m$ represent The temperature, humidity, light intensity and voltage data which collected by the same sensor node at the moment m respectively. Each element in the matrix equal to a pixel, and the value is used as the grayscale of the pixel in the matrix. Figure 1 is a gray image converted from IBRL data set after the above preprocessing. Figure 2, Figure 3, and Figure 4 are the corresponding gray-scale images after injecting a single fault into the temperature data in the IBRL data set. Figure 5, Figure 6 and Figure 7 convert the temperature data of IBRL data set into a corresponding gray image after injecting mixed faults into the IBRL data set respectively.

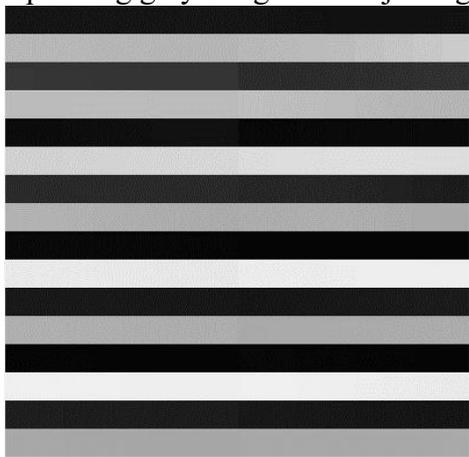

FIGURE 1. normal

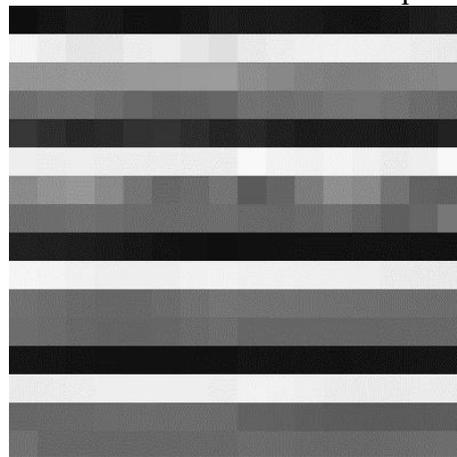

FIGURE 2. noise

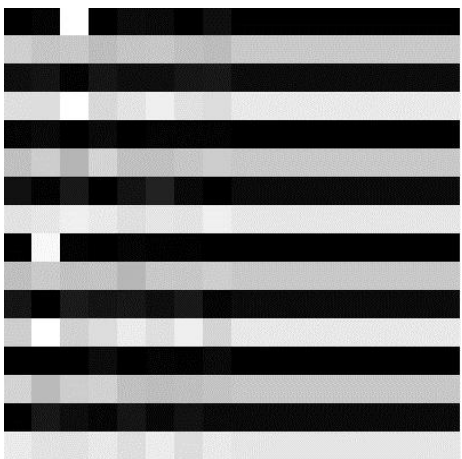

FIGURE 3. short

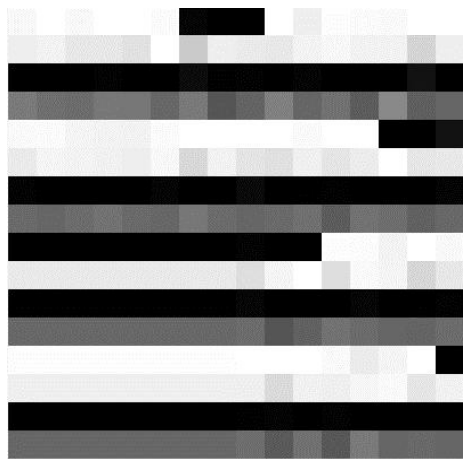

FIGURE 4. constant

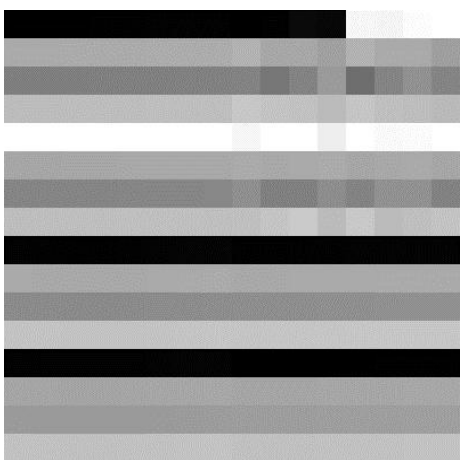

FIGURE 5. noise+constant

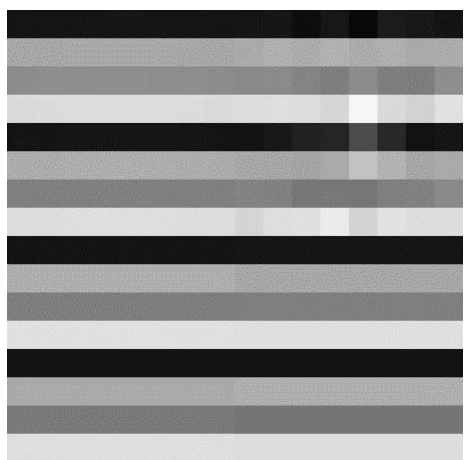

FIGURE 6. noise+short

— 5 —

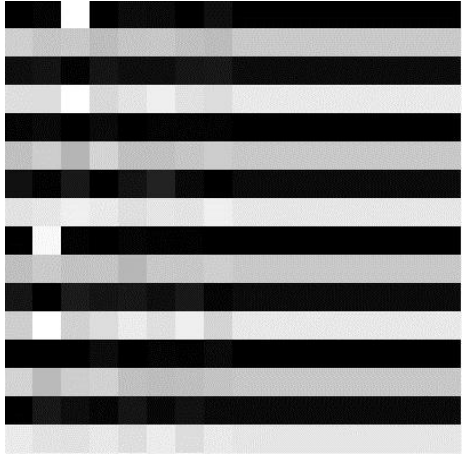
FIGURE 7.short+constant

1.3. **Structure design of convolutional neural network.** Input layer, convolution layer, pooling layer, full connection layer and output layer construct the base structure of Convolutional neural network （CNN）[16]。LeNet-5 is a classical convolutional neural network designed for handwritten digit recognition. On MNIST dataset, the LeNet-5 model can achieve an accuracy of about 99.4%. The LeNet-5 model has five convolution layers (three layers of convolution layer and two layers of full connection layer), plus two pooling layers, which make a total of seven layers of network. In this paper, we improve the traditional LeNet-5 model as follows. Firstly, the input layer of the network is designed as a 16 × 16 matrix because the data set is transformed into a grayscale image of appropriate size. Secondly, the purpose of data anomaly detection is to classify normal data and abnormal data, so the output layer of traditional LeNet-5 model is changed from 10 neurons to 2 neurons. Finally, three kinds of convolutional neural networks with different network structures are designed. Table 2 show the specific structures.

TABLE 2. Three kinds of convolutional neural networks with different structures

| Model | C1 | | S1 | | C2 | | S2 | | F1 |
|---|---|---|---|---|---|---|---|---|---|
| | Convolution kernel | Step size | Sampling window | Step size | Sampling window | Step size | Sampling window | Step size | Convolution kernel |
| M1 | 8×(3×3) | 1 | 2×2 | 2 | 16×(3×3) | 1 | 2×2 | 2 | 64×(4×4) |
| M2 | 8×(3×3) | 1 | 2×2 | 2 | 16×(5×5) | 1 | 2×2 | 2 | 64×(4×4) |
| M3 | 8×(3×3) | 1 | 2×2 | 2 | 16×(5×5) | 1 | 2×2 | 2 | 128×(4×4) |

The convolution neural network model designed in this paper is shown in Figure 8. The improved convolutional neural network based on LeNet-5 model has two convolution layers, two pooling layers and one full connection layer. C1 layer and C2 layer are convolution layers. In order to avoid the loss of image edge information and keep the same image size before and after convolution, C1 and C2 convolution layers are convoluted by all 0 filling. S1 layer and S2 layer are pool layer. One of the commonly used pooling layers is maximum pooling and the other is average pooling. S1 and S2 pool layers use the maximum pooling method to extract features. The full connection layer F1 transforms the two-dimensional feature map of S2 into one-dimensional vector, and takes the output result as the input of the classifier and output the classification result.

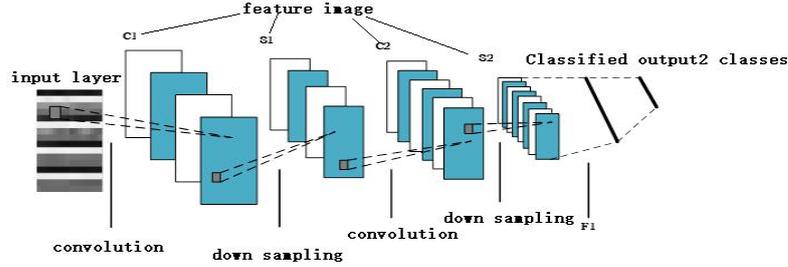

FIGURE 8. Convolutional neural network model

## 2. Experimental results and analysis of sampling window

2.1. **General requirements of experiment.** The text experiment environment is windows 10 which equipped 64 bit operating system, the model of graphics card is gtx1650 (4GB video memory), 8GB memory, and the processor is AMD ryzen5-2600. It is implemented by TensorFlow framework.

In order to verify the performance of the proposed algorithm, this paper uses DA (detection accuracy), TPR (true positive rate) and PRE (precision) as the performance indicators of the evaluation algorithm. Among them, DA is the detection accuracy rate, that is, the ratio of the number of correctly detected data in all the data participating in the detection to the total number of data participating in the detection. TPR is the correct detection rate. The number of abnormal data detected as abnormal data in the total. The ratio of the number of abnormal data, PRE is the detection accuracy.

$$DA = \frac{Number\ of\ samples\ tested\ correctly}{Number\ of\ samples\ tested} * 100\% \qquad (4)$$

$$TPR = \frac{Number\ of\ abnormal\ samples\ detected\ correctly}{Total\ number\ of\ abnormal\ samples} * 100\% \qquad (5)$$

$$PRE = \frac{Number\ of\ abnormal\ samples\ detected\ correctly}{Number\ of\ abnormal\ samples\ detected\ correctly + normal\ samples\ detected\ incorrectly} \qquad (6)$$

In this paper, only for the single fault and mixed fault of temperature data in IBRL data set. The experiment injects single fault and mixed fault into the temperature data collected by node 1 and node 2 in the IBRL data set .When we inject a mixed fault, r is set to 1.5, f is set to 1.5, and G is set to 300. In order to evaluate the detection performance of the proposed algorithm, the classical classification algorithm cart [10] is used to compare the performance of DA, TPR and pre in the case of single fault. Fig. 9, Fig. 10 and Fig show the specific experimental results under single fault. Table 3 show the specific experimental results under mixed fault.

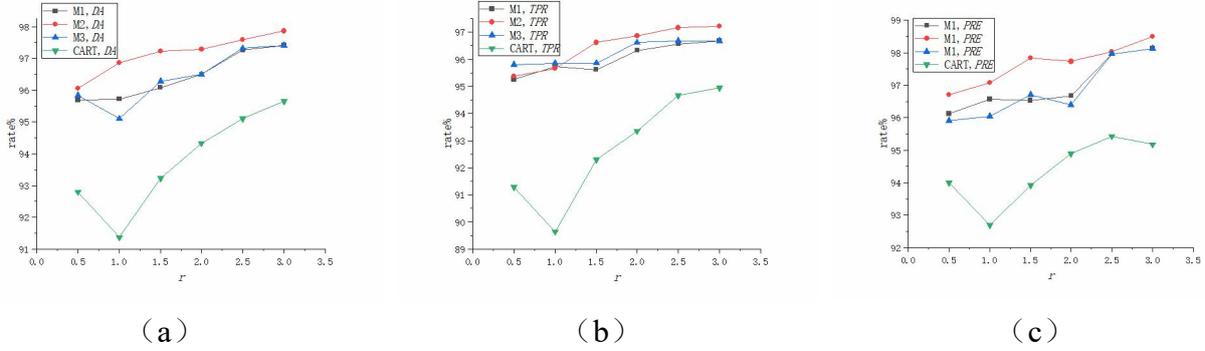

FIGURE 9. Three performance test results of four models under noise fault: (a) DA (b) TPR (c) PRE

From the test results in Fig. 9, it can be found that: 1) comparing the model M1 to M3 with cart, it is not difficult to find that the performance of model M1 to M3 is significantly higher than that of cart model in DA, TPR and PRE. We can concluded from Figures 9 (a) and 9 (c) that model M2 performs best in both DA and pre performance. From Figure 9 (b), we can observe that M2 performs best when R equals 1.5. To sum up, M2 is the optimal model in the process of R value increasing; 2) when R value gradually increases, the attribute values of DA, TPR and PRE of the four models are improved and gradually tend to be stable. This shows that for the three attributes, the increase of R value will improve the three performance of the model.

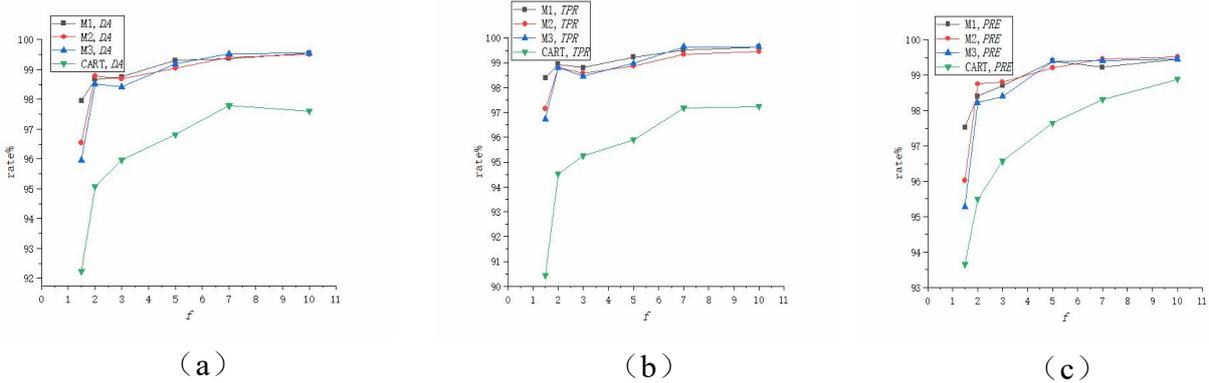

FIGURE 10. Three performance test results of four models under short fault: (a) *DA* (b) *TPR* (c) *PRE*

From the test results in Fig. 10, we can find that: 1) when the F value is the same, the performance of model M1 to M3 is significantly higher than cart model in the performance of DA, TPR and PRE. From figures 10 (a), (b) and (c), it is obvious that models M1 to M3 are consistent in DA, TPR and PRE basically, so the three models perform well. 2) When the value of F increases gradually, DA, TPR and PRE of the four models increase gradually and tend to be stable. From figures 10 (a), (b) and (c), it can be observed that after f is equal to value 5, models M1 to M3 tend to be stable gradually. However cart model is still in the growth trend, which also shows that the performance of M1 to m3 model is better than cart model.

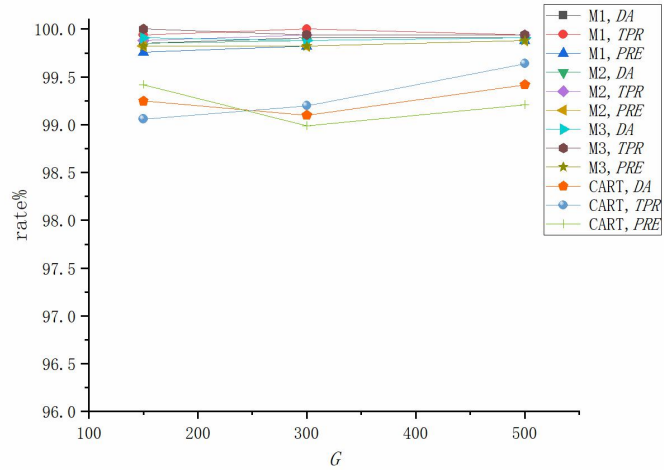

FIGURE 11. Three performance test results of four models under constant fault

From the test results in Fig. 11, we can conclude that: 1) among the four models, M1 to M3 models have strong advantages over cart model, and the DA, TPR and PRE of the three models are close to 100%. This shows that the performance of M1 to M3 is greatly improved compared with cart model. 2) When the value of G increases, the model M1 to M3 is stable basically, while the model cart has a small change, which indicates that the performance of model M1 to M3 is more stable.

TABLE 3. Experimental results of three models under mixed fault

| Mixed fault types | Model | DA | TPR | PRE |
|---|---|---|---|---|
| Noise+fixed | M1 | 99.94% | 100% | 99.88% |
| Noise+fixed | M2 | 99.94% | 100% | 99.88% |
| Noise+fixed | M3 | 99.94% | 100% | 99.88% |
| Noise+short | M1 | 97.15% | 96.86% | 97.44% |
| Noise+short | M2 | 97.06% | 97.16% | 96.98% |
| Noise+short | M3 | 96.70% | 96.92% | 96.52% |
| Short+fixed | M1 | 100% | 100% | 100% |
| Short+fixed | M2 | 99.90% | 100% | 99.88% |
| Short+fixed | M3 | 99.91% | 100% | 99.82% |

We can conclude some results in table 3.Firstly,in the case of noise and fixed mixed fault, the performance of the three models is consistent in all aspects.When using noise and short-term mixed fault, each of the three models has its own advantages and disadvantages. And there is no obvious difference in comprehensive performance.Meanwhile, when using short-term and fixed mixed fault, the three models perform very well without any difference.Generally speaking, model 2 performs better in three kinds of mixed faults. In addition, Compared with noise and fixed mixed fault, noise and short-term mixed fault, we can find that when there are noise faults, fixed fault injection can significantly improve the performance of the three models. Lastly,Comparing noise with fixed mixed fault, short-term and fixed mixed fault, it can be found that when there are fixed faults, the performance of the three models reaches the highest value. In this case, the detection algorithm is the best.

Through the above experiments, comprehensive analysis shows that the detection effect of injecting mixed fault is better than that of injecting single fault. In practical application, the fault type can not be determined in advance. After comprehensive comparison, model 2 will achieve better results in various situations obviously. Therefore, model 2 can be considered as the most adaptable and effective model among the three models.

## 3. Conclusion

This paper takes the data anomaly detection problem of wireless sensor network as the research object. On the basis of related research, the aspects of data preprocessing mode and convolutional neural network structure construct the convolutional neural network (CNN) model to realize the detection of abnormal data. In the experiment, we proposed three different network models and compared with the existing cart model, and the performance evaluation is carried out from the aspects of DA, TPR and PRE. The experimental results show that the three models proposed in this paper are better than cart model, and M2 model has the best performance. The research of deep learning algorithm in data anomaly detection is still in the immature stage. In the future research, I will focus on the correlation between the data collected by adjacent sensor nodes.